\documentclass[aps,pra,preprint,superscriptaddress,twocolumn,10pt]{revtex4-2}
\usepackage{natbib}
\usepackage[T1]{fontenc}
\usepackage[latin9]{inputenc}
\usepackage{graphicx,bm}
\usepackage{amsmath,amssymb}
\usepackage{color,xcolor,ulem,soul}

\usepackage[ bookmarks=true, colorlinks, linkcolor=blue, urlcolor=blue, citecolor=blue, plainpages=false, pdfpagelabels, final, breaklinks=true ]{hyperref}

\begin{document}


\title{Quantum heat engine in the optomechanical system with mechanical parametric drive}


\author{Zhen-Yang Peng}
\email[pengzhenyang@sjtu.edu.cn]{}

\affiliation{Wilczek Quantum Center, School of Physics and Astronomy, Shanghai Jiao Tong University, 200240, China}
\affiliation{CAS Key Laboratory of Theoretical Physics, Institute of Theoretical Physics, Chinese Academy of Sciences, Beijing 100190, China}
\author{Ying-Dan Wang}
\email[yingdan.wang@itp.ac.cn]{}
\affiliation{CAS Key Laboratory of Theoretical Physics, Institute of Theoretical Physics, Chinese Academy of Sciences, Beijing 100190, China}
\affiliation{School of Physical Sciences, University of Chinese Academy of Sciences, Beijing 100049, China}


\date{\today}
\begin{abstract}
We consider a quantum Otto-type heat engine constructed in an optomechanical system with which the cavity is chosen as the working substance. The cavity can effectively be coupled with hot thermal baths in nonequilibrium steady-states via optomechanical interaction. The mechanical mode with parametric drive fuels the cavity, and the utilization efficiency of energy is discussed. To obtain higher efficiency in finite time evolution, we use the shortcuts-to-adiabaticity method in work generation processes. The modified thermal efficiencies are obtained by numerical simulation. Such a system provides potential applications in quantum heat transfer and energy utilization in quantum devices.
\end{abstract}

\maketitle

\section{Introduction}

Quantum thermodynamics~\cite{quantumthermodynamics_review} is an extension of standard thermodynamics that the physical objects with the inclusion of quantum effects and usually in non-equilibrium situations. This rapidly growing research area not only provides the fundamental viewpoint on the understanding of quantum systems but also has great potential for nano-scale applications. As one of the important topics, quantum heat engines (QHEs) are important devices that convert heat to mechanical work for quantum systems~\cite{Scully_prl_2002, QuanHT_pre_2007, QHE_negativetemperature_prl, QHE_experiments_2019, QHE_flywheel_prl_2019}. Because of the quantum effects of the working substance or the heat baths, QHEs have unusual properties. For the cyclic heat engines to generate mechanical output work in a thermodynamic cycle, the efficiency of QHEs will surpass the standard limit defined in classical thermodynamics~\cite{Lutz_prl_2012, Lutz_prl_2014, QHE_negativetemperature_prl}. While for the steady-state heat engines that transfer energies from the hot baths to the cold baths, the QHE suggests that certain effects may appear to violate classical thermodynamic laws~\cite{Kosloff_epl_2014, Chiara_NJP_2018}. However, these effects are consistent with the principles of quantum mechanics.

For the fluctuation heat engines~\cite{fluctuationHE_review}, the probability distribution of work output~\cite{Lutz_pre_2007_workdistribution, Lutz_chemphys_2010} and heat absorbed~\cite{Lutz_pre_2018_heatdistribution} per thermodynamical cycle and the efficiency fluctuations~\cite{Broeck_NC_2014,Fazio_njp_2015,Lutz_prr_2020,MaYH_pra_2022} have been studied theoretically and experimentally. Because of the quantum non-adiabatic qualities, these distributions have unconventional behaviors, at the same time, the average work output and efficiencies will be suppressed. 

To design a quantum heat engine with non-zero power output, the finite-time thermodynamical processes will be taken into consideration for the optimization of the QHEs. The finite-time QHEs has been discussed for the finite-time adiabatic processes~\cite{SunCP_pre_2019, SunCP_pre_2019_2}, and by the technique of shortcut-to-adiabatically (STA)~\cite{Lutz_pre_2018_STA, STA_QHE_pre_2019, STA_review_RMP}. These two strategies lead to higher efficiencies and non-zero power.

In this paper, we have proposed a quantum harmonic heat engine constructed in an optomechanical system with an external mechanical parametric drive~\cite{Aash_njp_twophonon}. Unlike the other studies for QHEs in the optomechanical systems~\cite{ZhangKY_prl_OMHE, Zhangky_pra_2014, Dong_prl_2015, Lutz_prl_2014_nanomechanicalHE, Bathaee_pre_2016, Zhangky_pra_2016, Bennett_2020_fastOMHE, WuHB_sciadv_2021, Paulino_pra_2023}, the cavity is chosen as working substance. In particular, the mechanical mode has been considered as the quantum object that provides external energies, which is equivalent to the fuel as a component for classical machines. The heat flux for the total system in the cavity thermalization process is discussed, and the cyclic work extraction can be achieved from a single physical thermal reservoir coupled to the cavity. Due to the optomechanical interaction, non-equilibrium behavior is induced, resulting in this counter-intuitive phenomenon~\cite {Seifert_pre_2017}. We then construct the Otto-type cycle for the heat engine by modulating the parametric drive strength~\cite{mechanicalPD_experiment}. Finally, the extra heat absorption induced by the optomechanical interaction in the work generation strokes has been considered, and the corresponding thermal efficiencies with and without the STA method are analyzed.

This paper is organized as follows: in Sec.~\ref{sec: model} we first introduced the model, then the heat flux between the working fluid photons and the fuelled phonons is analyzed, and the quantum Otto cycle in this model is constructed; In Sec.~\ref{sec: efficiency} the efficiencies for the different cases are discussed. Conclusions and discussions are given in Sec.~\ref{sec: conclusion}. Some technical details are given in the Appendices.

\section{model and thermodynamic cycle} \label{sec: model}

We consider an optomechanical system with a parametric drive applied to mechanics~\cite{Aash_njp_twophonon, mechanicalPD_experiment} and a coherent laser drive $H_{\text{dri}}=A a e^{i\omega_L t}+ \text{h.c.}$, under the standard linearization, the system has the effective Hamiltonian
\begin{equation}
    \begin{split}
    H(t) &= H_a + H_b + G(a b^{\dagger}+a^{\dagger}b) \\
    &+ \frac{i}{2}\lambda (b^{\dagger 2}e^{-i2\omega_M t} - b^2 e^{i2\omega_M t}),\label{eq:systemH}
    \end{split}
\end{equation}
where $H_a = \Delta (a^{\dagger}a+1/2)$, $H_b = \omega_M (b^{\dagger}b+1/2)$, $\Delta$ is the cavity detuning, $\omega_M$ is the intrinsic frequency for mechanical mode, $G$ is the optomechanical coupling strength, $\lambda$ indicates the parametric drive strength. The cavity mode is chosen as the quantum harmonic heat engine, with the detuning change to produce work. Here the cavity detuning $\Delta=\omega_c-\omega_L$ can be tuned \textit{in situ}. In the weak interacting $G \ll \omega_M$ and resolved sideband regimes $\kappa \ll \omega_M$~\cite{Schliesser_2008_nphys, Teufel_2011_nature}, the optomechanical interaction would only be resonantly enhanced, and the dynamics of the total optomechanical system can be described by the following master equation:
\begin{equation}
\dot{\rho} = -i[H(t),\rho] + \mathcal{L}_{\text{diss}}\rho, \label{eq:mastereq}
\end{equation}
\begin{equation}
    \begin{split}
    \mathcal{L}_{\text{diss}}\rho &= \kappa (\bar{n}_c + 1) D[a]\rho + \kappa \bar{n}_c D[a^{\dagger}]\rho \\
    &+ \gamma (\bar{n} + 1)D[b]\rho + \gamma \bar{n} D[b^{\dagger}]\rho,
    \end{split}
\end{equation}
where $D[o]\rho=o\rho o^{\dagger}-(o^{\dagger}o \rho + \rho o^{\dagger}o)/2$. Based on the second law of thermodynamics, a heat engine must be in contact with at least two thermal baths, with different thermal occupations $\bar{n}_c$ and decay rate $\kappa$ in corresponding thermodynamic processes. However, when an auxiliary term that interacts with the heat engine is taken into account, this interaction introduces an additional excitation to the heat engine, effectively acting as a hot (or cold) reservoir that couples solely with the heat engine. In our work, the optomechanical interaction and the mechanical parametric drive would excite the cavity into higher occupations at the non-equilibrium steady-states, and thus produce an effective hot thermal bath for the heat engine. Therefore, the mechanical mode with parametric drive here fuels the cavity, and the interaction term transports the energy from the \textit{fuel} to the engine.

\subsection{Heat flux with the parametric drive}

As mentioned above, the mechanical mode acts as quantum fuel, providing effective hot reservoirs. Meanwhile, the cavity extracts heat from the mechanical mode via the optomechanical interaction. Naturally, the utilization efficiency of the energy of the quantum fuel, which can be defined as the ratio of heat flux transforming into the heat engine to the total heat flux injected into the mechanical mode, is one of the characteristics of quantum heat engines.

Based on energy conservation, the heat flux for both the cavity and the mechanical mode has the form~\cite{Kosloff_review(ARPC)}
\begin{subequations}
    \begin{align}
        \frac{d\langle H_a \rangle}{dt} &= \dot{Q}_{b \rightarrow a} + \dot{Q}^{\uparrow}_{a} + \dot{Q}^{\downarrow}_{a},\\
        \frac{d \langle H_b \rangle}{dt} &= \dot{Q}_{a \rightarrow b} + \dot{Q}_{\rm{PD}} + \dot{Q}^{\uparrow}_{b} + \dot{Q}^{\downarrow}_{b}.
    \end{align}\label{eq:heatflux}
\end{subequations}

\begin{figure}
    \centering
    \includegraphics[width=0.95\columnwidth]{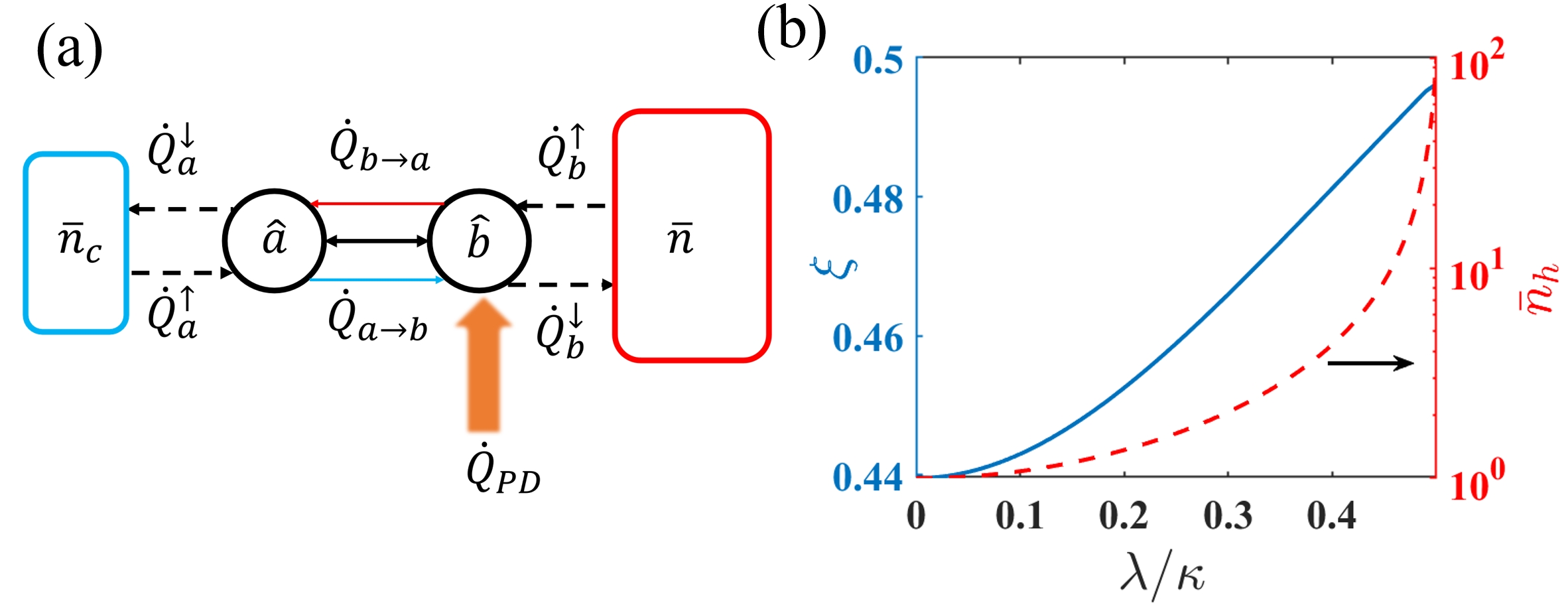}
    \caption{(a) Illustration for the heat flux and heat transfer between the cavity(heat engine) and the mechanical oscillator(quantum fuel) in the hot isochoric strokes. The mechanical parametric drive provides an external energy injection, while the optomechanical interaction transforms the energy from the "quantum fuel" to the heat engine. (b) Utilization efficiency and steady-state occupation numbers with parametric drive. Both the $\xi$ and the $\bar{n}_{h}$ are calculated under resonant case $\Delta=\omega_M$. Other parameters are chosen as $G=\kappa$, $\gamma=10^{-2}\kappa$, $\bar{n}_c=0.01$ and $\bar{n}=100$.}
    \label{fig:fig1}
\end{figure}
The detailed derivation is given in Appendix~\ref{sec:Appendix_1}, here $\dot{Q}_x$ indicates the different contribution for the heat flux. As illustrated in Fig.~\ref{fig:fig1}(a), $\dot{Q}_{\rm{PD}}$ is the energy injection via mechanical parametric drive, $\dot{Q}_{b \rightarrow a}$ indicates heat flux transferred from the fuel to the heat engine, $\dot{Q}^{\uparrow}_{i}(i=a,b)$ is the heating flux from the reservoir, $\dot{Q}^{\downarrow}_i{(i=a,b)}$ is the dissipating flux. The exact form of these heat flux terms can be derived from the master equation(Eq.~(\ref{eq:mastereq})),
\begin{subequations}
    \begin{align}
        \dot{Q}_{b \rightarrow a} &= i\omega_M G (\langle a^{\dagger}b \rangle - \langle ab^{\dagger} \rangle),\\
        \dot{Q}_{\rm{PD}} &= \omega_M \lambda (\langle b^{\dagger 2} \rangle + \langle b^2 \rangle),\\
        \dot{Q}^{\uparrow}_{b} &= \omega_M \gamma \bar{n} (\langle b^{\dagger}b\rangle +1),\\
        \dot{Q}^{\downarrow}_{b} &= -\omega_M \gamma(\bar{n}+1)\langle b^{\dagger}b \rangle.
    \end{align}\label{eq:heatfluxdef}
\end{subequations}
The definition of the utilization efficiency of energy for the mechanical mode is 
\begin{equation}
    \xi = \frac{-\dot{Q}_{b\rightarrow a}}{\dot{Q}_{\rm{PD}}+\dot{Q}^{\uparrow}_{b}},
\end{equation}
and the steady state occupation of photon mode is defined as $\bar{n}_h=\langle a^{\dagger}a \rangle_{ss}$~(Eq.~(\ref{eq:effnh})). The steady-state utilization efficiency of energy produced from the quantum fuel, and the effective thermal occupations for the cavity, are plotted in Fig.~\ref{fig:fig1}(b).
Compared with the standard optomechanical interaction case~($\lambda=0$), the external mechanical driven term generates additional energy injection, which supplies more energy than just extracted from the mechanical thermal bath. From the view of quantum thermodynamics, the larger steady-state occupation for the cavity indicates higher inner energy differences, which reflects more heat absorption or work production. On the other hand, following the discussion of heat flux for the total system, the larger parametric drive strength leads to a higher utilization efficiency of energy, thus more heat is transmitted from the mechanical mode into the cavity. Therefore, the cavity has a stronger ability to extract heat from the additional effective heat bath. Both of the above qualities indicate that in this optomechanical model, the cavity is a suitable candidate for heat engines used to produce mechanical output work, while the mechanical mode, which injects the additional heat into the engine and creates an effective hot thermal reservoir, has been acting as the quantum fuel.

\subsection{Quantum Otto-type Cycle}

A typical Otto cycle is constructed by two isochoric and two adiabatic processes. In the isochoric process, the engine is in contact with the hot(cold) thermal bath and has an energy exchange, while in the adiabatic process, the engine does not have thermal contact with the environment and just produces work. However, in quantum scenarios, the adiabatic processes not only satisfy the thermal adiabaticity that the system has zero thermal exchange but also follow the quantum adiabatic dynamics~\cite{QuanHT_pre_2007,Lutz_prl_2012, QHE_experiments_2019}. Therefore, we refer to the former case as the quantum Otto-type cycle.

By changing the mechanical parametric drive strength $\lambda$ in different steps, the quantum Otto-type cycle in our model, illustrated in Fig.~\ref{fig:fig2}, can be constructed as follows:

\begin{figure}
    \centering
    \includegraphics[width=0.95\columnwidth]{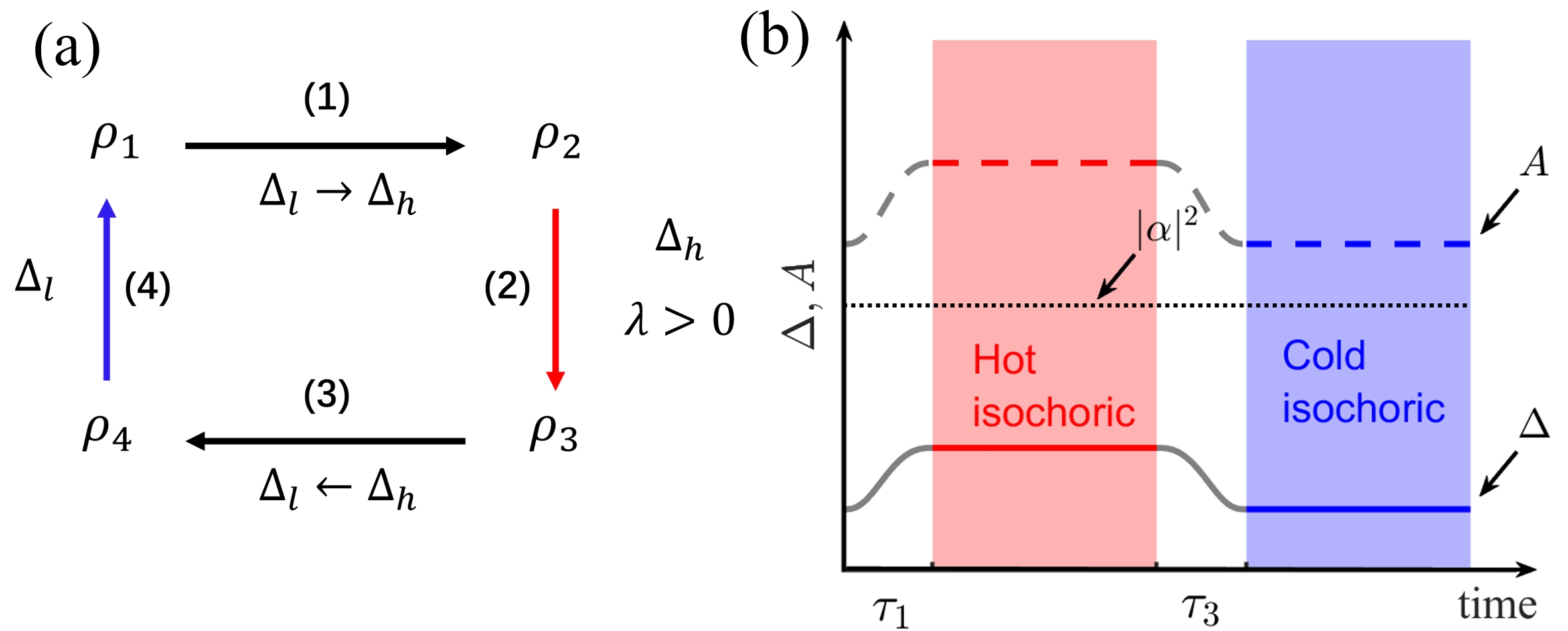}
    \caption{(a) The quantum Otto-type cycle for the cavity by tuning the detuning to produce the extra work output, and turning the parametric drive on and off to "heating" and "cooling" the cavity. (b) An illustration of the amplitude and detuning for the laser drive in an Otto-type cycle. The solid lines indicate detuning while the dashed lines are the corresponding amplitude. The black dashed-dotted line is $|\alpha|^2$, where $\alpha=A/(i\kappa-\Delta)$ is the classical part of the cavity mode. Thus the optomechanical coupling strength $G$ is fixed during the cycle because of $G\propto|\alpha|^2$.}
    \label{fig:fig2}
\end{figure}

(1). Isentropic compression($\rho_1\rightarrow\rho_2$), the cavity detuning $\Delta(t)$ and the amplitude of laser drive $A$ will be changed from ($\Delta_l$, $A_l$) to ($\Delta_h$, $A_h$) simultaneously during time $\tau$ to fix the classical part of the cavity mode. Thus in this stroke, the optomechanical coupling $G$ is fixed. During this stroke, $\lambda=0$.

(2). Hot isochoric($\rho_2\rightarrow\rho_3$), the optomechanical system with mechanical parametric drive $\lambda$, the cavity detuning $\Delta=\Delta_h=\omega_M$ is fixed. Then the system under the frame rotating $\omega_M$ is 
\begin{equation}
    H_h = G(ab^{\dagger}+a^{\dagger}b) + \frac{i}{2}\lambda (b^{\dagger 2}-b^2), \label{eq:hotHamiltonian}
\end{equation}
the steady-state occupation for the cavity mode after this stroke is $\bar{n}_h$. The system under the hot isochoric stroke has the stability condition that $\lambda \leq \min \{\frac{\gamma}{2}(1+C_0),\frac{\gamma+\kappa}{2} \}$, where $C_0=4G^2/\kappa \gamma$.

(3). Isentropic expansion($\rho_3\rightarrow\rho_4$), during this stroke the cavity detuning will reverse to $\Delta_l$ via $\Delta(\tau-t)$, the heat engine will produce positive work.

(4). Cold isochoric($\rho_4\rightarrow\rho_1$), the system evolves to the steady state without mechanical parametric drive ($\lambda=0$), the cavity detuning is fixed at $\Delta_l$.

To design the two isentropic strokes, the evolution time $\tau \ll 1/\kappa$ is chosen to neglect the energy dissipation. Hence, in the large detuned case, $|\Delta_l-\Delta_h| \gg \kappa $, the steady state for the cavity in the cold isochoric stroke would almost be the thermal state with average occupation $\bar{n}_c$, and the optomechanical interaction term can be regarded as perturbation in analyzing the two isentropic strokes. Thus, the thermal efficiency in the absence of optomechanical interaction during the two isentropic processes can serve as a benchmark. Meanwhile, while the mechanical parametric amplification during the hot isochoric stroke does indeed generate squeezing for the cavity field, the steady state remains a Gaussian state (see Appendix~\ref{sec:Appendix_2} for the exact form of the steady-state P-function). Consequently, the thermal decomposition of a Gaussian state~\cite{Gaussian_Quantum_Information_RMP} indicates that, after the hot isochoric stroke, the eigenstate of the cavity mode follows a thermal distribution with an average occupation number $\bar{n}_h$.

The work production and heat absorption in a quantum Otto cycle can be defined with the two-point measurement approach, in which the energy change for the heat engine is obtained by projective measurement at the beginning and the end of each stroke~\cite{nonequilibriumFR_RMP}. The joint probability distribution for an Otto cycle has the form~\cite{Lutz_prr_2020}
\begin{equation}
    \begin{split}
    &P(W,Q) = \sum_{m,n,j,k} \delta \left[ W -(E^l_j-E^h_k+E^h_m-E^l_n) \right] \\
    &\times \delta \left[ Q-(E^h_k-E^h_m) \right] P^{\tau}_{n\rightarrow m} P^{\tau}_{k \rightarrow j} P_n(\bar{n}_c) P_j(\bar{n}_h).
    \end{split}
\end{equation}
Here $P^{\tau}_{n\rightarrow m}$ and $P^{\tau}_{k \rightarrow j}$ indicate the transition probabilities from initial eigenstates $n(k)$ to the final eigenstates $m(j)$ for the isentropic strokes~\cite{Husimi_1953,Lutz_pre_2007_workdistribution,Lutz_chemphys_2010}, and $P_n(N)=(1/(1+N))(N/(1+N))^n$ is the probability distribution for the thermal states with occupation $N$. 

\begin{figure}
    \centering
    \includegraphics[width=0.95\columnwidth]{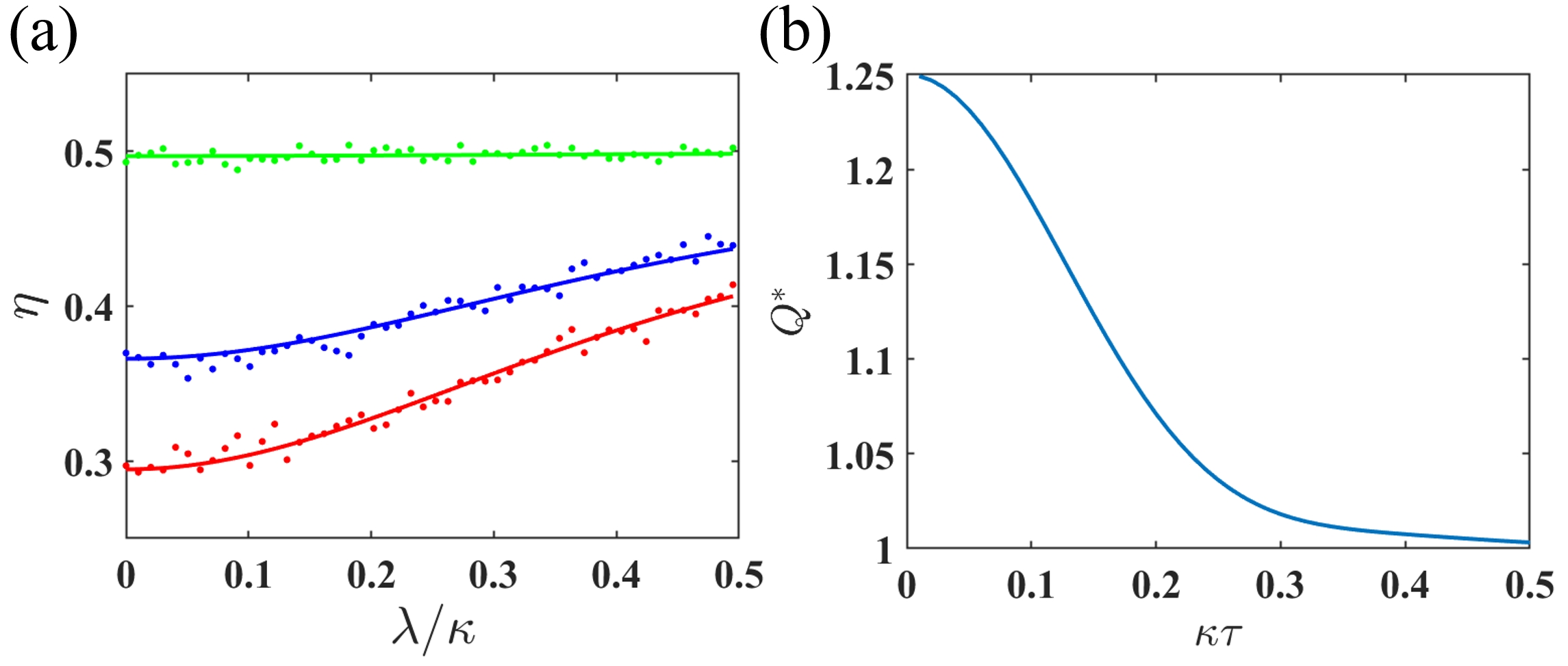}
    \caption{(a) The non-adiabatic thermal efficiencies~(Eq.~(\ref{eq:eta})). The red, blue, and green lines correspond to the chosen evolution time $\tau = 0.1/\kappa$, $0.15/\kappa$ and $0.5/\kappa$, respectively. The solid lines are the analytical results, while the dots are the numerical simulation results. (b) The adiabatic parameter under different $\tau$. In both plots, $Q^*$ is calculated with the detuning as Eq.~(\ref{eq:delta_t_form}), other parameters are chosen as $\Delta_h = 20\kappa$, $\Delta_l=\Delta_h/2$, $\omega_M=\Delta_h$, $\bar{n}=100$, $\bar{n}_c=0.01$ and $\gamma=0.01\kappa$. }
    \label{fig:fig3}
\end{figure}

For the calculation of average net work output and heat absorbed without optomechanical interaction in the isentropic strokes per cycle, the characteristic function of the joint probability distribution is
\begin{widetext}
\begin{equation}
    \chi(u,v) = \int dW dQ e^{iuW} e^{ivQ} P(W,Q) = \chi_h(u,v) \chi_l(u,v),
\end{equation}
\begin{equation}
    \begin{aligned}
    \chi_l(u,v) &= \frac{\sqrt{2}}{\bar{n}_c+1} \{ Q^* (1-e^{i2(u-v)\Delta_h}) [ 1-e^{-i2u\Delta_l}( \frac{\bar{n}_c}{\bar{n}_c+1})^2 ] \\
    &+ (1+e^{i2(u-v)\Delta_h}) [ 1+e^{-i2u\Delta_l} ( \frac{\bar{n}_c}{\bar{n}_c+1} )^2 ] -4e^{i(u-v)\Delta_h}e^{-iu\Delta_l}( \frac{\bar{n}_c}{\bar{n}_c+1} ) \}^{-1/2},
    \end{aligned}
\end{equation}
\begin{equation}
    \begin{aligned}
    \chi_h(u,v) &= \frac{\sqrt{2}}{\bar{n}_h+1} \{ Q^* (1-e^{i2u\Delta_l}) [ 1-e^{-i2(u-v)\Delta_h}( \frac{\bar{n}_h}{\bar{n}_h+1})^2 ] \\
    &+ (1+e^{i2u\Delta_l}) [ 1+e^{-i2(u-v)\Delta_h} ( \frac{\bar{n}_h}{\bar{n}_h+1} )^2 ] -4e^{-i(u-v)\Delta_h}e^{iu\Delta_l}( \frac{\bar{n}_h}{\bar{n}_h+1} ) \}^{-1/2}.
    \end{aligned}
\end{equation}
\end{widetext}
 The average work and heat absorbed from the effective hot reservoir are obtained from the characteristic function~\cite{MaYH_pra_2022}
 \begin{subequations}
    \begin{align}
    \langle W \rangle_{Q^*} &= -i \frac{\partial \text{ln}\chi(u,v)}{\partial u} \mid_{u=v=0},\\
    \langle Q \rangle_{Q^*} &= -i \frac{\partial \text{ln}\chi(u,v)}{\partial v} \mid_{u=v=0}.
    \end{align}
\end{subequations}
The thermal efficiency can be derived as
\begin{equation}
    \eta_{\text{th}} = 1 - \frac{\Delta_l}{\Delta_h} \frac{Q^*(\bar{n}_h+1/2)-(\bar{n}_c+1/2)}{(\bar{n}_h+1/2)-Q^*(\bar{n}_c+1/2)}.\label{eq:eta}
\end{equation}
Here $Q^*$ is the adiabatic parameter, for $Q^*=1$ is the quantum adiabatic condition while $Q^*>1$ the non-adiabatic scattering occurs~\cite{Husimi_1953}. Fig.~\ref{fig:fig3} indicates that the dissipation can be safely neglected when the evolution time in the two isentropic processes is small. 

\section{Work generation and efficiencies} \label{sec: efficiency}

So far the heat absorption in a quantum Otto-type cycle is discussed. To describe quantum heat engines, work generation is another important process to determine thermal efficiency. Below we will discuss the two strokes in the quantum Otto-type cycle that generate work.

The quantum Otto engine reaches its maximal thermal efficiency when the isentropic strokes evolve quantum adiabaticity($Q^*=1$), with which the adiabatic Otto efficiency $\eta_{\rm{Qtto}}=1-\Delta_l/\Delta_h$. However, the quantum adiabatic condition requires the dynamical evolution of the system to be slow enough that the non-adiabatic transitions between different eigenstates can be neglected, and the system follows the instantaneous eigenstates in time. Therefore, even if the quantum adiabatic processes would have maximal efficiency, the power for the total work generation is zero.  To obtain the efficiency close to the adiabatic case in finite time, the STA technique~\cite{STA_review_RMP, Lutz_pre_2018_STA} has been considered.

Here we use the counter-diabatic driving(CD) method to realize STA. The basic method is to find a driving Hamiltonian, with which the exact solution for the total Hamiltonian including the driving term, is the adiabatic approximating result for the original Hamiltonian~\cite{quantum_sys_fremod}. The CD Hamiltonian has the form~\cite{Lutz_pre_2018_STA}
\begin{equation}
    H_{\rm{STA}}(t) = i \frac{\dot{\Delta}_t}{4\Delta_t} (a^2 - a^{\dagger 2}).\label{eq:H_STA}
\end{equation}

The STA method has boundary conditions, which ensures the CD term vanishes at the initial and the final time,
\begin{equation}
    \begin{split}
        \Delta_i = \Delta_{l(h)}, \ \Delta_f = \Delta_{h(l)},\\
        \dot{\Delta}_{t=0}=\dot{\Delta}_{t=\tau}=0, \\
         \ddot{\Delta}_{t=0}=\ddot{\Delta}_{t=\tau}=0.
    \end{split} \label{eq:bondcon}
\end{equation}
As an example, we can choose the detuning in compression strokes~\cite{Lutz_pre_2018_STA}
\begin{equation}
    \begin{split}
    &\Delta_t = \Delta_i + 10(\Delta_f - \Delta_i)(t/\tau)^3 \\
    &-15(\Delta_f - \Delta_i)(t/\tau)^4 + 6(\Delta_f - \Delta_i)(t/\tau)^5, \label{eq:delta_t_form}
    \end{split}
\end{equation}
where $\Delta_i=\Delta_l$, $\Delta_f=\Delta_h$. The expansion strokes have the detuning $\Delta_{(\tau-t)}$ with opposite initial and final values compared with compression strokes.

As we have analyzed in Sec.~\ref{sec: model}, the optomechanical interaction induces an effective hot thermal bath for the heat engine. Thus, the additional CD term changes the thermal occupation for the cavity, which indicates additional heat absorption during compression/expansion strokes. The average value for the corresponding process is $\langle H^{c/e}_{\rm{STA}}\rangle_{\tau} = \frac{1}{\tau}\int^{\tau}_0 \text{d}t \langle H^{c/e}_{\rm{STA}}(t) \rangle$, where the instantaneous expectation value can be obtained
\begin{equation}
    \langle H^{c/e}_{\rm{STA}}(t)\rangle = \frac{\Delta_{t/(\tau-t)}}{\Delta_i} \langle H(0) \rangle \left( \frac{\Delta_{t/(\tau-t)}}{\Omega_{t/(\tau-t)}} -1 \right),
\end{equation}
here the superscript c/e indicates the compression/expansion processes, and $\Omega_{t/(\tau-t)} = \Delta_{t/(\tau-t)} \sqrt{1-\dot{\Delta}^2_{t/(\tau-t)}/4\Delta^4_{t/(\tau-t)}}$ \cite{Yuki_pre_transprob} is the modified eigen-frequency for the total system that includes the CD term. Thus in the STA evolution, the parameters should be chosen to keep $\Omega_{t/(\tau-t)}>0$, and the modified efficiency has the form
\begin{equation}
    \eta_{\rm{STA}} = - \frac{\langle W \rangle_{Q^*=1}}{\langle Q \rangle_{Q^*=1} +\sum_{j=c,e}\langle H^{j}_{\rm{STA}}\rangle_{\tau}}. \label{eq:eta_STA}
\end{equation}

\subsection{Extra heat produced by optomechanical interaction}

So far we have discussed the quantum Otto heat engine whose cavity acts as the working fluid in an optomechanical system with mechanical parametric drive by neglecting the optomechanical interacting term in the two isentropic strokes. Hence, in the compression/expansion processes, the optomechanical system is equivalent to two independent harmonic oscillators, the time-dependent cavity detuning only produces work. However, the optomechanical interaction can not be turned on and off straightforwardly, even weak coupling changes the energy for the cavity. Therefore, the external interaction leads to extra heat transfer between the cavity and the mechanical mode.

There are a few cases in which the OM interaction can be neglected in the work-producing processes, e.g., the detuning suddenly changes from initial to final values. In this case, the adiabatic parameter $Q^*=(\Delta_i^2+\Delta_f^2)/2\Delta_i \Delta_f$~\cite{Husimi_1953,Lutz_prl_2012} is larger than 1 and the STA process can not be realized. To realize STA, the time-dependent detuning (Eq.(\ref{eq:delta_t_form})) needs finite evolution time to satisfy the boundary condition~(Eq.~(\ref{eq:bondcon})), thus it leads to the extra heat transfer from the mechanical mode via the optomechanical interaction. In our quantum heat engine model, the mechanical mode acts as the quantum fuel, which provides an effective hot thermal reservoir to the cavity. Thus during the compression/expansion processes, the optomechanical interaction generates extra heat absorption.

The extra heat absorbed from the mechanical mode at time $t$ can be written as
\begin{equation}
    \begin{split}
    \Delta E^{c/e}(t) &= E^{\text{int},c/e}_a (t) - E^{\text{0},c/e}_a (t),\\
    E^{\text{int/0},c/e}_a(t) &= \langle H_a (t)\rangle_{\text{int/0}} = \text{Tr} \{ \rho^{\text{int/0}} H_a (t) \}.
    \end{split}
\end{equation}
The superscript $\text{int/0}$ indicates within/without optomechanical interaction, and the $\rho^{\text{int/0}}$ corresponds to
\begin{subequations}
\begin{align}
\dot{\rho}^{\text{int}} &= -\text{i}[H_a+H_b+H_{\text{int}},\rho^{\text{int}}] + \mathcal{L}_{\text{diss}}\rho^{\text{int}},\\
\dot{\rho}^0 &= -\text{i}[H_a+H_b,\rho^0]+\mathcal{L}_{\text{diss}}\rho^0,
\end{align}
\end{subequations}
where the $H_{\text{int}}$ indicates the optomechanical interaction term in Eq.~(\ref{eq:systemH}).Similarly, the average heat absorption during compression/expansion processes is
\begin{equation}
    \langle \Delta E^{\rm{c/e}} \rangle_{\tau} = \frac{1}{\tau} \int^{\tau}_0 \text{d}t \Delta E^{\rm{c/e}}(t).\label{eq:heat_OM}
\end{equation}
Thus the modified efficiency is 
\begin{equation}
    \eta^{\prime}_{\rm{STA}} = - \frac{\langle W \rangle_{Q^*=1}}{\langle Q \rangle_{Q^*=1} +\sum_{j=c,e} \langle H^{j}_{\rm{STA}}\rangle_{\tau} + \langle \Delta E^{j} \rangle_{\tau}}.\label{eq:eta_prime_STA}
\end{equation}

\begin{figure}
    \centering
    \includegraphics[width=0.95\columnwidth]{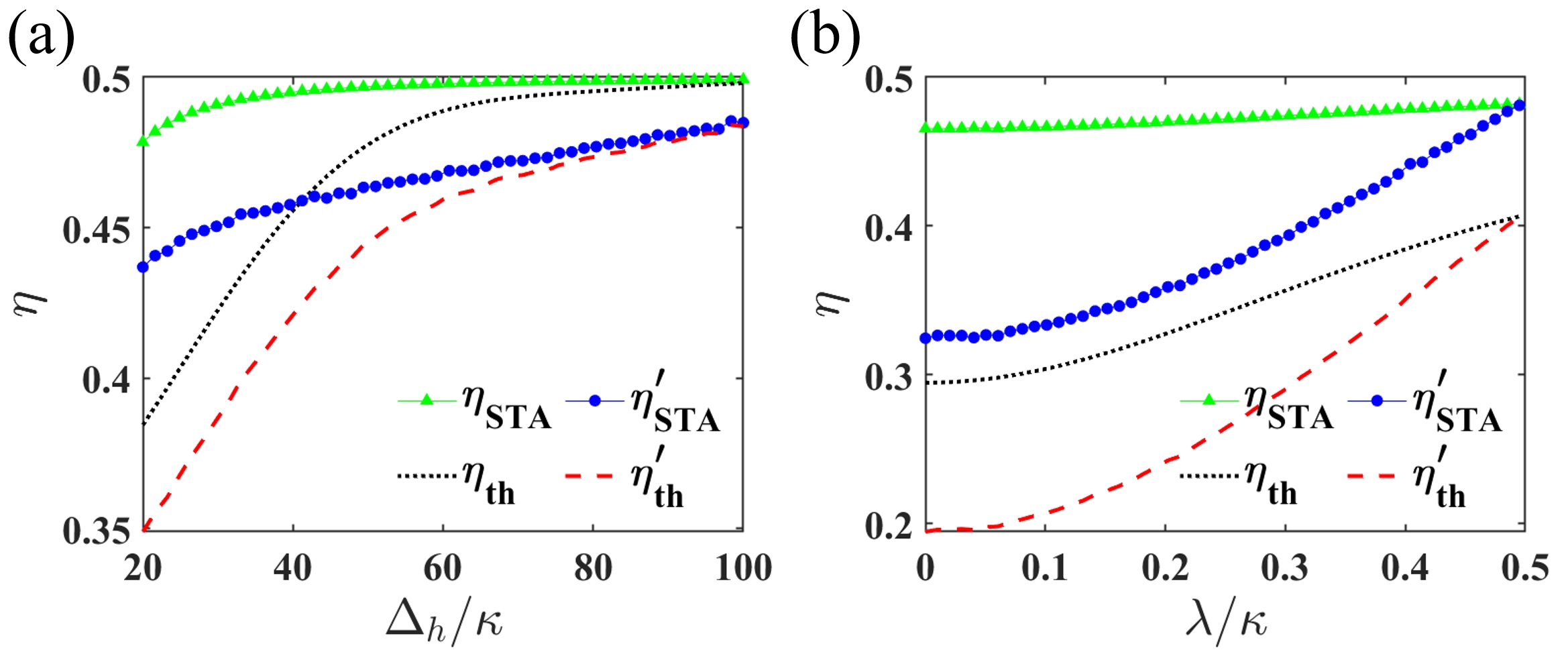}
    \caption{Efficiencies for different controlling parameters under fixed $\tau$. (a) Efficiencies for different detunings with a fixed ratio of $\Delta_l/\Delta_h=1/2$, where the mechanical parametric driving strength $\lambda=0.4\kappa$; (b) Efficiencies vs $\lambda$ with the same ratio of $\Delta_l/\Delta_h=1/2$ and fixed $\Delta_h=20\kappa$. In both plots, the evolution time has been chosen as $\kappa \tau = 0.1$. Other parameters for both plots are chosen as $\bar{n}_c=0.01$, $\bar{n}=100$, $G=\kappa$, $\gamma=0.01\kappa$ and $\omega_M=\Delta_h$.}
    \label{fig:fig4}
\end{figure}

The thermal efficiency, when optomechanical coupling is taken into account, is given by $\eta^{\prime}_{\text{th}}=-\langle W \rangle_{Q^*}/(\langle Q\rangle_{Q^*}+\sum_{j=c,e}\langle \Delta E^j \rangle_{\tau})$. Here $Q^*>1$ because of the nonadiabatic nature, and the additional heat absorption $\langle \Delta E^{\rm{c/e}} \rangle_{\tau}$, resulting from the optomechanical interaction, has been included.
The efficiencies for the fixed evolution time $\tau$ are plotted in Fig.~\ref{fig:fig4}. In both plots, the nonadiabatic efficiencies are calculated by Eq.~(\ref{eq:eta}), the modification for the heat absorption induced by the optomechanical interaction is added to the total heat absorption, and the detuning was tuned following the Eq.~(\ref{eq:delta_t_form}). The efficiencies for the STA methods that neglect and include the extra heat terms are calculated by Eq.(\ref{eq:eta_STA}) and Eq.~(\ref{eq:eta_prime_STA}), respectively. It has been noted that as $\Delta_h$ and $\Delta_l$ increase, the STA Hamiltonian(Eq.~(\ref{eq:H_STA})) has less contribution to the system evolution, the system evolves near the instantaneous eigenstates once the boundary conditions are fulfilled. Meanwhile, because the contribution to the total heat absorption in a thermodynamic cycle has three parts, heat absorbing from the effective hot thermal reservoir $\langle Q \rangle$, the counter-diabatic driving $\langle H^{c/e}_{\rm{STA}}\rangle_{\tau}$ and the extra heat from the optomechanical interaction(Eq.~(\ref{eq:heat_OM})), the contribution for the latter two terms is mainly determined by the evolution time for the two strokes, thus the influence of these two parts decreases as mechanical parametric driving strength increases.

\subsection{Optimal Efficiency Control}

Compared to the traditional thermodynamic cycle, a finite-time heat engine explicitly accounts for finite-time cycles, which generate a non-zero power output. To obtain the higher output power, the evolution time in working processes is required as short as possible. In our model, due to the limit condition $\Omega_t>0$ to avoid the trap inversion~\cite{Lutz_pre_2018_STA}, there exists the shortest evolution time in compression/expansion strokes. Below this limit, $\Omega_t$ becomes a complex value, and the STA path breaks down. 

\begin{figure}
    \centering
    \includegraphics[width=0.95\columnwidth]{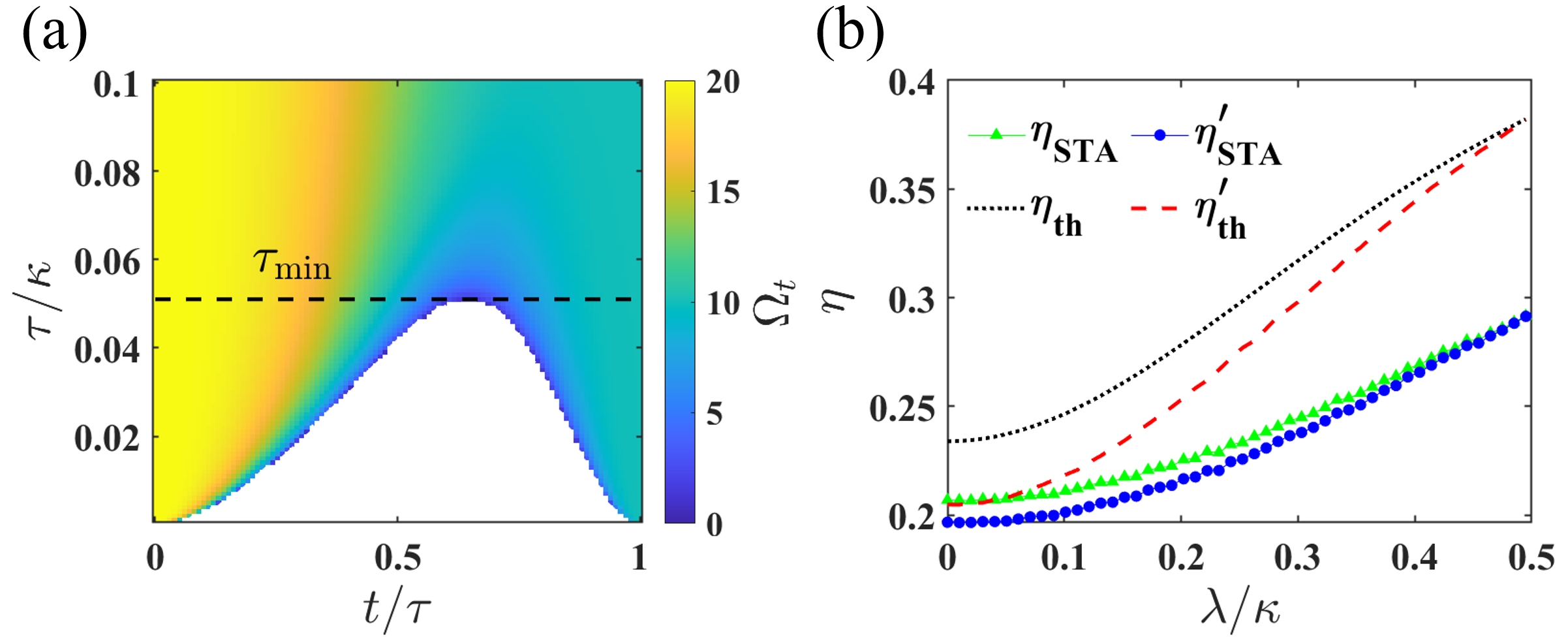}
    \caption{(a) The minimal $\tau$ in the compression process, which is obtained for $\Omega_t >0$, the white area in the plot means trap inversion occurs. (b) Efficiency at minimal $\tau$ vs $\lambda$. The corresponding $\tau_{\text{min}}$ is obtained numerically. Other parameters are chosen the same as Fig.~\ref{fig:fig4}.}
    \label{fig:fig5}
\end{figure}

Fig.~\ref{fig:fig5}(a) illustrates the minimal evolution time $\tau_{\text{min}}$ in the working strokes. The corresponding efficiencies for the heat engine under the $\tau_{\text{min}}$ are plotted in Fig.~\ref{fig:fig5}(b), with the same corresponding efficiencies as in Fig.~\ref{fig:fig4}. It has been noticed that under the $\tau_{\text{min}}$ in the two working strokes, $\eta_{\text{STA}}<\eta_{\text{th}}$. For a time-dependent system, the faster time evolution usually has a larger non-adiabatic scattering rate, thus, the CD term needs higher extra energy to limit the system evolution close to the adiabatic eigenstates, the heat exchange induced by the STA path $\langle H^{c/e}_{\text{STA}} \rangle_{\tau}$, has a significant influence on the thermal efficiencies. Therefore, although the working processes have the largest output power, the STA path in $\tau_{\text{min}}$ leads to an efficiency even smaller than the non-adiabatic thermal efficiency.

\begin{figure}
    \centering
    \includegraphics[width=0.95\columnwidth]{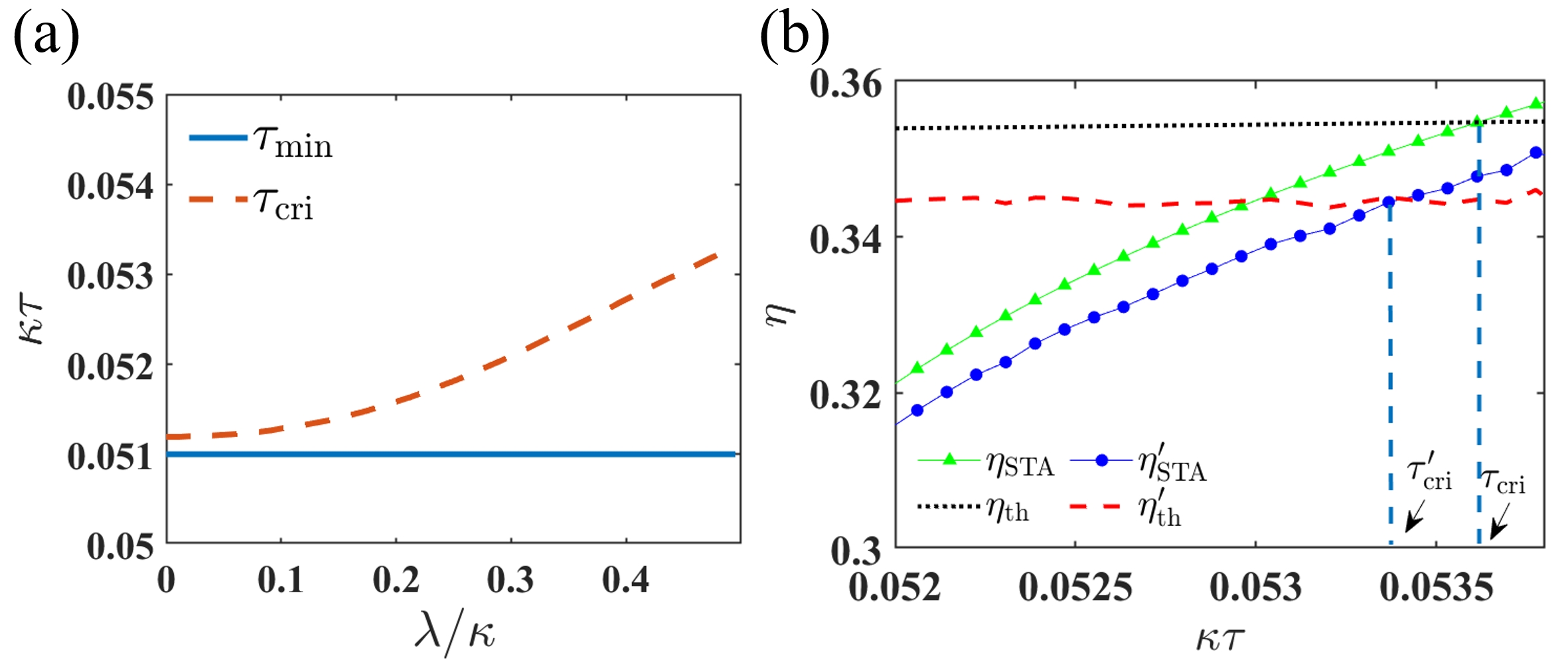}
    \caption{(a) Minimal and critical evolution time. (b) Efficiencies vs $\tau$. The dashed vertical lines indicate the critical time with ($\tau^{\prime}_{\text{cri}}$) and without ($\tau_{\text{cri}}$) optomechanical coupling. It can be noticed that the optomechanical interaction decreases the critical time. Other parameters are chosen the same as Fig.~\ref{fig:fig4}.   }
    \label{fig:fig6}
\end{figure}

Due to the $\tau$-dependency for the additional heat absorption $\langle H^{c/e}_{\rm{STA}}\rangle_{\tau}$ generated in working strokes, it can be deduced that there exists a critical evolution time $\tau_{\text{cri}}$ to distinguish whether the STA efficiency (Eq.~(\ref{eq:eta_STA})) is larger than the non-adiabatic thermal efficiency (Eq.~(\ref{eq:eta})) or not, and for $\tau \lessgtr \tau_{\text{cri}}$, $\eta_{\text{th}} \gtrless \eta_{\text{SAT}}$. Under $\tau_{\text{cri}}$ that without optomechanical coupling, $\eta_{\text{th}}=\eta_{\text{STA}}$, which suggests
\begin{equation}
    \sum_{j=c,e} \langle H^j_{\text{STA}} \rangle_{\tau} = \frac{\langle W \rangle_{Q^*=1} \langle Q \rangle_{Q^*}}{\langle W \rangle_{Q^*}} - \langle Q \rangle_{Q^*=1}. \label{eq:cri_condition}
\end{equation}
The critical time $\tau_{\text{cri}}$ is defined to maintain the critical condition (Eq.~(\ref{eq:cri_condition})) that the contribution of heat absorption of the STA counteracts the non-adiabatic effect, which increases when $\lambda$ increases, as plotted in Fig.~\ref{fig:fig6}(a). Once the optomechanical coupling has been considered in the working strokes, the extra heat exchange $\langle \Delta E^{c/e}\rangle_{\tau}$ induced by the interaction leads to the extra heat absorption to the engine, with which $\sum_{j=c,e}\langle \Delta E^j\rangle_{\tau}>0$. This extra heat absorption requires more energy for the STA path to suppress the non-adiabatic scattering. Therefore, for the critical time $\tau^{\prime}_{\text{cri}}$ (corresponding to $\eta^{\prime}_{\text{STA}} = \eta^{\prime}_{\text{th}}$) when optomechanical coupling is taken into account, $\tau^{\prime}_{\text{cri}} < \tau_{\text{cri}}$, as illustrated in Fig.~\ref{fig:fig6}(b).

\section{conclusions and discussions} \label{sec: conclusion}

We proposed a quantum heat engine model in a standard optomechanical system with an additional mechanical parametric drive term. We find the non-equilibrium steady-state exhibits an amplification for cavity excitation, which is equivalent to an effective thermal distribution with high thermal occupations for the photon mode. A cyclic quantum Otto-type heat engine that the cavity as working fluid is discussed, and the engine in contact with effective hot and cold thermal baths can be realized by switching the mechanical parametric drive on and off. We also discussed heat extraction from the auxiliary quantum system, which can be designed as quantum fuel. The utilization efficiency for the quantum fuel is defined. In our model, the thermal efficiency and the utilization efficiency of energy have been enhanced simultaneously via the additional parametric mechanical drive. Our study might be useful for later research on energy transformation and supplements in quantum devices.

\begin{acknowledgements}
Y.D.W. acknowledges support from the Fundamental and Interdisciplinary Frontier Research Priority Program of Chines Academy of Sciences (Grant No. XDB0920102), NSFC (Grant No.12275331) and the Penghuanwu Innovative Research Center (Grant No. 12047503).
\end{acknowledgements}

\appendix
\section{Heat flux}\label{sec:Appendix_1}

The equations for the heat flux (Eq.~(\ref{eq:heatflux}, \ref{eq:heatfluxdef})) in the main text can be derived from the master equation (Eq.~(\ref{eq:mastereq})). Using the equation $\partial_t\langle O \rangle = \text{Tr}\{\dot{\rho}O\}$, we have
\begin{equation}
\begin{split}
\frac{\text{d}\langle H_a \rangle}{\text{d}t} &= -i \text{Tr}\{[H_{\text{int}},\rho]H_a\} \\
&+ \kappa (\bar{n}_c + 1)\text{Tr}\{D[a]\rho H_a\} + \kappa \bar{n}_c \text{Tr}\{D[a^{\dagger}]\rho H_a\},
\end{split}
\end{equation}
\begin{equation}
\begin{split}
\frac{\text{d}\langle H_b \rangle}{\text{d}t}
&= -i\text{Tr}\{[H_{\text{int}},\rho]H_b\} -i\text{Tr}\{[H_{\text{PD}},\rho] H_b\} \\
&+ \gamma(\bar{n}+1)\text{Tr}\{D[b]\rho H_b\} + \gamma \bar{n} \text{Tr}\{D[b^{\dagger}]\rho H_b\}.
\end{split}
\end{equation}
Here the $H_{\text{int}}$ and $H_{\text{PD}}$ correspond to the interaction and mechanical parametric drive terms, respectively. Each part of the above equations indicates the energy transformation between two subsystems $H_a$ and $H_b$, as illustrated in Fig.~\ref{fig:fig1}(a). If we define 
\begin{subequations}
\begin{align}
\dot{Q}_{a(b)\rightarrow b(a)} &= -i \text{Tr}\{ [H_{\text{int}},\rho]H_{b(a)}\},\\
\dot{Q}_{\text{PD}} &= -i\text{Tr}\{[H_{\text{PD}},\rho]H_b\},\\
\dot{Q}^{\uparrow}_{a} &= \kappa \bar{n}_c\text{Tr}\{D[a^{\dagger}]\rho H_a\},\\
\dot{Q}^{\downarrow}_{a} &= \kappa (\bar{n}_c +1)\text{Tr}\{D[a]\rho H_a\},\\
\dot{Q}^{\uparrow}_{b} &= \gamma \bar{n} \text{Tr}\{D[b^{\dagger}]\rho H_b\},\\
\dot{Q}^{\downarrow}_{b} &= \gamma (\bar{n}+1)\text{Tr}\{D[b]\rho H_b\},
\end{align}
\end{subequations}
then the exact form of Eq.~(\ref{eq:heatflux},\ref{eq:heatfluxdef}) can be obtained.

\section{The Fokker-Planck Equation}\label{sec:Appendix_2}

The master equation (Eq.~(\ref{eq:mastereq})) with the hot isochoric Hamiltonian (Eq.(\ref{eq:hotHamiltonian})) can be easily transformed into the phase space, with the positive-P function Fokker-Planck equation form
\begin{equation}
    \frac{\partial P(\vec{\alpha},t)}{\partial t} = \left( -\vec{\alpha}^{\prime T} A \vec{\alpha} + \frac{1}{2}\vec{\alpha}^{\prime T} D \vec{\alpha}^{\prime} \right) P(\vec{\alpha},t), \label{eq: FP}
\end{equation}
where $\vec{\alpha}=(\alpha, \ \alpha^*, \ \beta, \ \beta^*)^T$, $\vec{\alpha}^{\prime}=(\partial/\partial \alpha, \ \partial/\partial \alpha^*, \ \partial/\partial \beta, \ \partial/\partial \beta^*)^T$,
\begin{gather}
    A = \begin{pmatrix}
        -\kappa/2 & 0 & -i G & 0\\
        0 & -\kappa/2 & 0 & i G \\
        -i G & 0 & -\gamma/2 & \lambda\\
        0 & i G & \lambda & -\gamma/2
    \end{pmatrix}
\end{gather}
\begin{gather}
    D = \begin{pmatrix}
        0 & \kappa \bar{n}_c & 0 & 0\\
        \kappa \bar{n}_c & 0 & 0 & 0\\
        0 & 0 & \lambda & \gamma \bar{n}\\
        0 & 0 & \gamma \bar{n} & \lambda
    \end{pmatrix}.
\end{gather}
The matrix $A$ can be diagonalized as $\tilde{A}=SAS^{-1}=\rm{diag}$$(k_1, \ k_2, \ k_3, \ k_4)$, where
\begin{equation*}
    \begin{split}
        k_{1(2)}&=\frac{1}{4}\sqrt{-\gamma-\kappa-2\lambda \mp \sqrt{(\gamma-\kappa+2\lambda)^2-16G^2} }\\
        k_{3(4)}&=\frac{1}{4}\sqrt{-\gamma-\kappa+2\lambda \mp \sqrt{(\kappa-\gamma+2\lambda)^2-16G^2} }.
    \end{split}
\end{equation*}
To let all the above four eigenvalues $k_i$ have negative real parts, we can finally get the stability condition $\lambda \leq \min \{\gamma(1+C_0)/2,(\gamma+\kappa)/2\}$.

Following the Green's function method, we obtained the steady state solution of Eq.~(\ref{eq: FP})
\begin{equation}
    P_{ss}(\vec{\alpha})=\frac{1}{\sqrt{(2\pi)^4 \det |Q_{ss}|}}\exp \left(-\frac{1}{2} \vec{\alpha}^T Q^{-1}_{ss} \vec{\alpha} \right), \label{eq:ssP_fun}
\end{equation}
with $Q_{ss}=S^{-1}\tilde{Q}_{ss}(S^{-1})^T$, $(\tilde{Q}_{ss})_{ij}=-\tilde{D}_{ij}/(k_i+k_j)$ and $\tilde{D}=SDS^T$. Then the steady-state correlations  can be solved
\begin{equation}
    \langle \bar{O}_i O_j \rangle_{ss} = (Q_{ss})_{ij}/2,
\end{equation}
where $\vec{O}=(a,\ a^{\dagger},\ b,\ b^{\dagger})^T$. Therefore, the steady-state photon occupation and correlations which reflected the utilization efficiency of energy have the form
\begin{equation}
    \begin{split}
    \langle a^{\dagger}a \rangle_{ss} &= \frac{1}{2} \left. \big \{ \frac{\kappa \bar{n}_c(\gamma-2\lambda)(\gamma+\kappa-2\lambda)+4G^2(\gamma\bar{n}+\kappa \bar{n}_c+\lambda)}{[4G^2+\kappa(\gamma-2\lambda)](\gamma+\kappa-2\lambda)} \right.\\
    &+ \left.\frac{\kappa \bar{n}_c(\gamma+2\lambda)(\gamma+\kappa+2\lambda)+4G^2(\gamma\bar{n}+\kappa \bar{n}_c-\lambda)}{[4G^2+\kappa(\gamma+2\lambda)](\gamma+\kappa+2\lambda)} \right. \big \},
    \end{split}\label{eq:effnh}
\end{equation}
\begin{equation}
    \begin{split}
        \langle b^{\dagger}b \rangle_{ss} &= \frac{1}{2} \left. \big \{ \frac{\kappa(\gamma+\kappa-2\lambda)(\bar{n}\gamma+\lambda)+4G^2(\gamma\bar{n}+\kappa \bar{n}_c+\lambda)}{[4G^2+\kappa(\gamma-2\lambda)](\gamma+\kappa-2\lambda)} \right.\\
        &+ \left. \frac{\kappa(\gamma+\kappa+2\lambda)(\bar{n}\gamma-\lambda)+4G^2(\gamma\bar{n}+\kappa \bar{n}_c-\lambda)}{[4G^2+\kappa(\gamma+2\lambda)](\gamma+\kappa+2\lambda)} \right. \big \},
    \end{split}
\end{equation}
\begin{equation}
    \begin{split}
        \langle a^{\dagger}b \rangle_{ss} &= iG\kappa \left. \big \{ \frac{(\bar{n}-\bar{n}_c)\gamma+(1+2\bar{n}_c)\lambda}{[4G^2+\kappa(\gamma-2\lambda)](\gamma+\kappa-2\lambda)} \right.\\
        &+ \left. \frac{(\bar{n}-\bar{n}_c)\gamma-(1+2\bar{n}_c)\lambda}{[4G^2+\kappa(\gamma+2\lambda)](\gamma+\kappa+2\lambda)} \right. \big \},
    \end{split}
\end{equation}
\begin{equation}
    \begin{split}
        \langle b^{2} \rangle_{ss} &= \left. \big\{ \frac{\kappa(\gamma+\kappa-2\lambda)(\bar{n}\gamma+\lambda)+4G^2(\gamma\bar{n}+\kappa \bar{n}_c+\lambda)}{[4G^2+\kappa(\gamma-2\lambda)](\gamma+\kappa-2\lambda)} \right.\\
        &- \left. \frac{\kappa(\gamma+\kappa+2\lambda)(\bar{n}\gamma-\lambda)+4G^2(\gamma\bar{n}+\kappa \bar{n}_c-\lambda)}{[4G^2+\kappa(\gamma+2\lambda)](\gamma+\kappa+2\lambda)} \right. \big\}.
    \end{split}
\end{equation}

\section{stochastic differential equations}

The above Fokker-Planck equation (Eq.~\ref{eq: FP}) in different scenarios is equivalent to the Ito stochastic differential equation, which has the form
\begin{equation}
    \text{d} \vec{\bm{\alpha}}_{t}^{\text{int/0,c/e}} = A(\vec{\bm{\alpha}}_{t}^{\text{int/0,c/e}}) \text{d}t + B(\vec{\bm{\alpha}}_{t}^{\text{c/e}})\text{d}\vec{W}_t,
\end{equation}
where $\vec{\bm{\alpha}}_t$ is the column vector formed from the families of random variables $\vec{\alpha}_{1t}, \dots \vec{\alpha}_{nt}$, each $n$ indicates a group of independent trajectories in phase space; $A(\vec{\bm{\alpha}}_t^{\text{int/0,c/e}})$ is the drift term, while the matrix $B(\bm{\alpha}_t^{\text{int/0,c/e}})$ is defined by the diffusion matrix:
\begin{equation}
    D(\bm{\alpha}) = B(\bm{\alpha}) B(\bm{\alpha})^T.
\end{equation}
And $\vec{W}_t$ indicates n independent Wiener process, which in the short time limit is the term of Gaussian white noise. The matrix $A^{\text{int/0,c/e}}$,$B^{\text{c/e}}$ have the form
\begin{widetext}
\begin{gather}
    A^{\text{int,c/e}} = \begin{pmatrix}
    -\kappa/2-i\Delta_{t/(\tau-t)} & 0 & -iG & 0 \\
    0 & -\kappa/2+i\Delta_{t/(\tau-t)} & 0 & iG \\
    -iG & 0 & -\gamma/2-i\omega_M & 0 \\
    0 & iG & 0 &-\gamma/2+i\omega_M
    \end{pmatrix},
\end{gather}
\begin{gather}
    A^{\text{0,c/e}} = \begin{pmatrix}
    -\kappa/2-i\Delta_{t/(\tau-t)} & 0 & 0 & 0 \\
    0 & -\kappa/2+i\Delta_{t/(\tau-t)} & 0 & 0 \\
    0 & 0 & -\gamma/2-i\omega_M & 0 \\
    0 & 0 & 0 &-\gamma/2+i\omega_M
    \end{pmatrix},
\end{gather}
\end{widetext}
\begin{gather}
    B^{\text{c/e}} = \frac{1}{\sqrt{2}}\begin{pmatrix}
        i \sqrt{\kappa \bar{n}_{h/c}} & \sqrt{\kappa \bar{n}_{h/c}} & 0 & 0 \\
        -i \sqrt{\kappa \bar{n}_{h/c}} & \sqrt{\kappa \bar{n}_{h/c}} & 0 & 0 \\
        0 & 0 & i \sqrt{\gamma \bar{n}} & \sqrt{\gamma \bar{n}} \\
        0 & 0 & -i \sqrt{\gamma \bar{n}} & \sqrt{\gamma \bar{n}}
    \end{pmatrix}
\end{gather}

Thus the instantaneous inner energy for the cavity can be calculated as 
\begin{equation}
    E^{\text{int/0,c/e}}_{\text{cav}}(t) = \Delta_{t/(\tau-t)}(\langle \bm{\alpha}^{\text{int/0}}_1 \bm{\alpha}^{\text{int/0}}_2 \rangle +1/2).
\end{equation}

\bibliography{ref_QHE,new}

\end{document}